\documentstyle[epsf]{l-aa}
\begin{document}

 \thesaurus{05         
              (05.01.1;03.13.2;04.03.1;10.15.2 M67)}  

\title{
A new approach to the reduction of  ``Carte du Ciel'' plates}

\author {A. Ortiz-Gil\inst{1} \thanks{{\em Present address: Dept. of
Astrophysics \& Optics, School of Physics,
The University of New South Wales, Sydney 2052, NSW, Australia} 
} \and M. Hiesgen\inst{2} \and P. Brosche \inst{1}}
\offprints{A. Ortiz-Gil} \institute{ Sternwarte der Universit\"at
Bonn, Auf dem H\"ugel 71, D-53121 Bonn, Germany \and
Observatoire de Strasbourg, 11 rue de l'Universit\'e, 67000 Strasbourg, France
} \date{Received 20 May 1997\ ; accepted 4 August 1997\ }

\maketitle

\begin{abstract}
A new procedure for the reduction of ``Carte du Ciel'' plates is presented.
A typical ``Carte du Ciel'' plate corresponding to the Bordeaux zone has been
taken as an example. It shows triple exposures for each object and the 
modeling of the data has been performed by means of a non-linear least 
squares 
fitting of the sum of three bivariate Gaussian distributions. A number of
solutions for the problems present in this kind of plates (optical 
aberrations, adjacency photographic effects, presence of grid lines, emulsion
saturation) have been investigated. An internal accuracy of $0^{''}\!\!.1$ in 
$x$ and $y$ was obtained for the position of each of the individual exposures. 
The external reduction to a catalogue led to results with an accuracy 
of $0^{''}\!\!.16$ in $x$ and $0^{''}\!\!.13$ in $y$ for the mean position of 
the three exposures. A photometric calibration 
has also been performed and magnitudes were determined with an accuracy of 
$0^{\mbox{\scriptsize m}}\!\!.09$ .

\end{abstract}

\section{Introduction}

The ``Carte du Ciel'' (CdC in what follows)
was one of the first large international joint astronomical projects of the 
history.
It was officially initiated in 1887 with a double target: the construction
of a sky catalogue complete down to the $11^{\mbox{\footnotesize th}}$ 
magnitude (which eventually became the Astrographic Catalogue) and the
construction of a sky chart down to the $14^{\mbox{\footnotesize th}}$ 
magnitude. 
These projects represent the very first photographic registers of the 
whole sky.  In consequence, they constitute a very important source of data
for proper motion studies (see, for example, Geffert et al. 1996)
because of the large time interval when comparing with today observations 
(about one hundred years in most of the cases) as well as for studies on 
galactic structure and kinematics.

Part of the data was completely measured and the results were published as 
``Astrographic Catalogue''. A re-reduction with these published data has been 
recently performed by Urban \& Corbin (1996). Other groups are taking
a complete CdC second epoch plates (Potter et al. 1996).
On the other hand, Carte plates
have never been systematically measured because they were not intended to give 
positions but only to be a sky chart. But due to the deeper images taken 
(we have identified stars fainter than $B=15$ 
(Hiesgen et al. in preparation) )  these plates are of great importance 
because of the large amount of data they contain and the time elapsed since 
they were produced. 

\begin{figure}
\def\epsfsize#1#2{.35\hsize}
\centerline{\epsffile{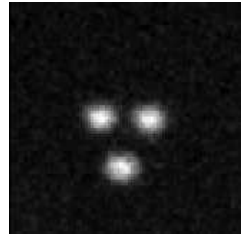}}
\caption[]{One star as example for the triple exposures on the CdC plates}
\label{fig7}
\end{figure}

The plates were exposed three times, with the exposures being placed
at the vertices of an approximately equilateral triangle (see Fig.~\ref{fig7}).
They were also provided with a grid of perpendicular lines in order to ease 
the star position measurements. These two characteristics make the accurate 
measurement of positions very difficult in some specific cases because of:
1) merging of the three images when the star is brighter than
$B \approx 12 $ and then the presence of an adjacency photographic
effect called ``Kostinsky effect'' (Kostinsky 1907; Ross 1921) which 
increases the measured distance
between the exposures, and 2) stars lying on a grid line or close enough
to let the Kostinsky effect show up. Optical aberrations (mainly
spherical aberration and field curvature) are also present.

These problems
have prevented astrometrists till now from using these triple image
plates as a valuable source of data, although there are some punctual
exceptions (see Geffert et al. 1996, for a recent example). In this paper
our aim is to show that good astrometric and photometric results are 
achievable, following the general procedure outlined in this work.
A different procedure was proposed by Bonnefond (1991) and used by Geffert
et al. (1996). Our method is able to deal with stars brighter than the upper 
limit of $V=9.5$ in Geffert et al. (1996) because it can work with 
blended images. 

A European Community Human and Capital Mobility
Network was initiated in 1994 under the name of ``Salvaging an Astrometric
Treasure'' (Ortiz-Gil et al. 1995, Hiesgen et al. 1996) with the scope of 
measuring 
these plates and using the data for a variety of applications, such as 
extension of the Hipparcos reference system to fainter magnitudes, 
determination of proper motions with an accuracy of $2 \times 10^{-3}$ 
arsec/yr, proper motions 
of globular and open clusters, photographic magnitudes with an accuracy of 
about $0.1 $, or studies related to galactic structure and kinematics.

Different typical CdC plates have been analysed in order to study how to 
deal with the 
specific problems which arise and may vary much from plate to plate. Therefore,
a complete study of the characteristics of each individual plate is necessary. 
Here we present the complete analysis 
performed with a
plate which was gently provided by Bordeaux Observatory (France). It was
taken in February 1922 and covers a 2$^{\mbox{\scriptsize o}}$x2$^{\mbox{\scriptsize o}}$ field 
with centre at $\alpha= 8^{\mbox{\scriptsize h}}50^{\mbox{\scriptsize m}}$ and $\delta = 11^{\mbox{\scriptsize o}}
 48^{'}$ (1900 equinox). This field is interesting because it includes the 
open  cluster M\,67. It shows three exposures (Fig.~\ref{fig7}), placed 
at the vertices of a triangle. 

The plate was scanned with the PDS 2020 GM$^{plus}$ microdensitometer 
of the Astronomical Institute M\"unster (Germany). In the process of scanning
the plate on-line search and segmentation techniques were employed, which 
produced an output file containing detected objects in the form of 
``picture frames'' of individual images (Horstmann 1988). 
These frames have a size of $61 \times 61$~pixels ($0.61\times0.61$~mm), or 
$81 \times 81$~pixels ($0.81\times0.81$~mm) in the case of bright stars, to 
assure that all three exposures belonging to the same triangle are included 
in the frame. 

Visual inspection of these
frames allows us to remove from the set those frames in which any of the three
exposures is missing, there is confusion between exposures due to the 
presence of double stars or the system has identified a spurious 
image. This implies that we can end up with at most three different frames 
corresponding to the same triplet. They are identified by their equal 
positions on the plate and the one which will be finally used in the study 
of the plate characteristics and reduction will be the one giving a better 
fit to the triple Gaussian model described later.

In the rest of this work we have assigned the following numbers
to the three exposures: exposure 1 (top left exposure in Fig.~\ref{fig7}), 
exposure 2 (top right exposure in Fig.~\ref{fig7}) and exposure 3 (bottom 
exposure in Fig.~\ref{fig7}).

\section{Modeling the data}

To calculate the positions of each of the components of these triple images 
the fitting with a model consisting on the sum of three bivariate Gaussian
distributions was applied (Dick et al. 1993). The density $D_{i,j}$,
measured in pixel $i,j$, is then:

\small
\begin{eqnarray}
\begin{array}{l l}
D_{ij}= & B + \sum\limits_{k=1}^{3} A_{k} 
\exp \left\{ - \frac{1}{2} \left[ \frac{1}
{1-t_{k}^{2}} 
\left( \left( \frac{x_{ij}- x_{ck}}{\sigma_{xk}} \right)^{2} \right. \right. 
\right. \\
& 
+ \left. \left. \left. \left( \frac{y_{ij}-y_{ck}}{\sigma_{yk}} \right)^{2} 
-  2t_{k} \left( \frac{x_{ij}-x_{ck}}
{\sigma_{xk}} \right) \left( \frac{y_{ij}-y_{ck}}{\sigma_{yk}} \right) 
\right) \right]^{s_k} \right\} 
\end{array}
\end{eqnarray}

\normalsize
\rm
In this expression $B$ is the background of the frame. As the triple images are
confined to areas of $0.81 \times 0.81$~mm at most, $B$ can be considered 
constant in each of these frames (see Sect. \ref{pca}). $A_{k}$ is the peak 
density of the $k$-th image, while
$x_{ij},y_{ij}$ are the pixel coordinates and $x_{ck},y_{ck}$ are the centre
coordinates of the $k$-th image. $\sigma_{xk},\sigma_{yk}$ and $t_k$ are 
the parameters 
of an ellipse with arbitrary orientation and axis size.
These three parameters are related to the ellipse semimajor and semiminor
axis, $a_k$ and $b_k$, by the expressions:

\small
\begin{equation}
a_{k}= \left\{ \frac{1}{2} \left[ (\sigma^2_{xk}+\sigma^2_{yk}) + 
\sqrt{(\sigma^2_{xk}- \sigma^2_{yk})^{2}+
4t_{k}^{2} \sigma^2_{xk} \sigma^2_{yk}} 
\right] \right\} ^{ \frac{1}{2}}  
\label{eq2}
\end{equation}
\begin{equation}
b_{k}=\left\{ \frac{1}{2}\left[(\sigma^2_{xk}+\sigma^2_{yk}) - 
\sqrt{(\sigma^2_{xk}-\sigma^2_{yk})^{2}+4t_{k}^{2}
\sigma^2_{xk}\sigma^2_{yk}} \right] \right\}^{\frac{1}{2}}  
\label{eq3}
\end{equation}

\noindent
{\normalsize and the angle between X and $a_k$ axes is}

\small
\begin{equation}
\theta_{k}= arcos \frac{1}{\sqrt{1+ \left( \frac{\lambda_{k}-\sigma^2_{xk}}
{t_k \sigma_{xk} \sigma_{yk}} \right)^{2}}}
\label{eq4}
\end{equation}

\noindent
{\normalsize where}

\small
\begin{equation}
\lambda_{k}= \frac{1}{2}\left[ (\sigma^2_{xk}+\sigma^2_{yk}) + 
\sqrt{(\sigma^2_{xk}-\sigma^2_{yk})^{2}+4t_{k}^{2}
\sigma^2_{xk}\sigma^2_{yk}} \right]
\label{eq5}
\end{equation}

\normalsize

The parameter $s_k$ is a flattening parameter which takes into account the 
sa\-tu\-ra\-tion of the photographic emulsion.

The fitting process was performed by means of the Levenberg-Marquardt method
(Marquardt 1963) for non-linear least squares fitting to work out the values of the 22 free
parameters in the model. The fitting was applied to small frames of 
$0.61 \times 0.61$ mm ($61 \times 61$ pixels), or $0.81 \times 0.81$ mm 
($81 \times 81$ pixels) for bright stars, around each set of triple images. 
One of our main objectives is to reduce the number of these free parameters 
in order to achieve better results (see Sect. \ref{pca}). At the moment this 
software has not been fully prepared for its use by other groups, but we 
intend to make it available to the community as soon as possible.

\section{Optical aberrations}

Four optical aberrations would mainly be expected to show up in an optical 
system of the kind used in the CdC astrographs. These
are coma, spherical aberration, field curvature and chromatic aberration.
In particular, coma makes the images to become asymmetric as it gives them
a comet-like appearance. 
Spherical aberration and field curvature generate extended
and elongated images, respectively, from pointlike ones.
How much elongated and asymmetric the images are is of importance
in order to see how the optical aberrations are influencing the final 
results in the plate reduction.  

\subsection{Ellipticity}

Images at a radial distance from the plate centre greater than $\approx 5$\ cm
show a significant ellipticity, which is defined as:\\

\begin{equation}
e_k=1- \frac{b_k}{a_k}
\label{eq6}
\end{equation}

\noindent
where $a_k,b_k$ are the image semimajor and semiminor axes of exposure $k$, 
respectively, given by Eqs. (\ref{eq2}) and (\ref{eq3}).

\begin{figure}
\def\epsfsize#1#2{.65\hsize}
\centerline{\epsffile{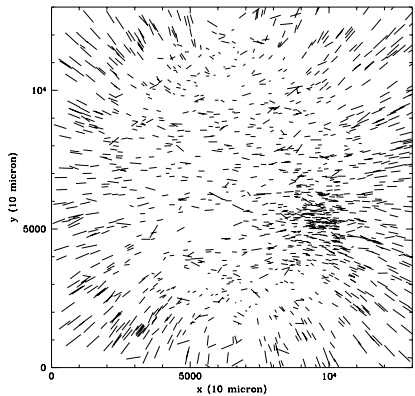}}
\caption{Field curvature in the plate. The length of the
lines is proportional to the mean ellipticity of the three exposures of
each stellar image. The orientation with respect to the $x$-axis is 
proportional to the mean of the three angles $\theta_{k}$}
\label{fig2}
\end{figure}

\begin{figure}
\def\epsfsize#1#2{.65\hsize}
\centerline{\epsffile{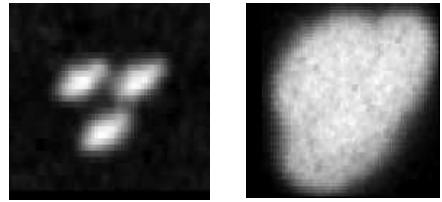}}
\caption[]{Effects of optical aberrations are shown for two stars of different 
magnitudes}
\label{fig5}
\end{figure}

In Fig.~\ref{fig2} the ellipticity as a function of the plate position is
shown. Segment lengths are proportional to the mean ellipticity of the three
exposures, while their orientation with respect to the X axis is
the mean of the three angles $\theta_{k}$. We expect to encounter this 
effect in all CdC plates because they all were taken with
a similar kind of optics. Two examples of these images are shown in 
Fig.~\ref{fig5}. 

\begin{figure}
\def\epsfsize#1#2{.75 \hsize}
\centerline{\epsffile{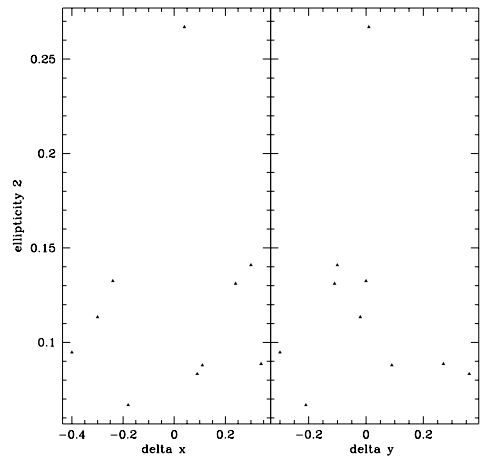}}
\caption[]{Image ellipticities as a function of the
residuals in $x$ and $y$ for the three exposures after performing
the plate reduction with an external catalogue. No dependence on the 
ellipticity value is evident here.
Ellipticity values of exposure 2 have been used as an example.
Analog diagrams are obtained with the ellipticity values of exposures 1 and 3} 
\label{fig13}
\end{figure}

No significant dependence of the position error on
the ellipticity can be found after performing the plate reduction with the 
external catalogue (Fig.~\ref{fig13}). This suggests that the model is 
therefore able to deal correctly with elongated images. 

\subsection{Image asymmetry}

A study on the asymmetry of the images in the plate has been carried out by 
means of analysing the skewness of the marginal distribution along the 
``possible'' asymmetry axis:

\begin{equation}
skew=\frac{\sum_{i} f(x_i)(x_i-\bar{x})^3 / \sum_{i} f(x_i)}
{\sigma^3} 
\end{equation}

\noindent
where $f(x_i)$ is the marginal distribution along the image semimajor axis
and $\sigma$ stands for the marginal distribution standard deviation.

\begin{figure}
\def\epsfsize#1#2{.75 \hsize}
\centerline{\epsffile{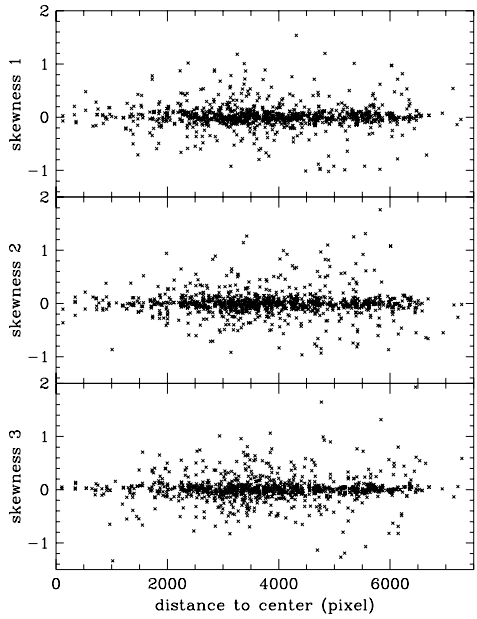}}
\caption[]{Skewness values as a function of the distance to the plate centre.
$88.7\%$ of the values lie within [-0.2,0.2]}
\label{fig15}
\end{figure}

Most of the skewness values ($88.7\%$) lie on the interval [-0.2,0.2]
(Fig.~\ref{fig15}), which
seems to indicate that the images can be considered symmetric and,
in consequence, that coma is not the dominant cause for the distortion
but rather the field curvature and spherical aberration. The last
one makes the pointlike images to appear as disks while the first one
is responsible for the elongation of these disks.

\begin{figure}
\def\epsfsize#1#2{.65 \hsize}
\centerline{\epsffile{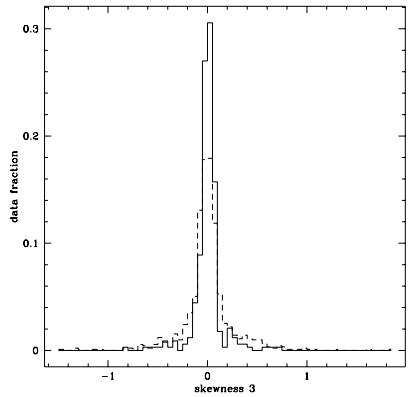}}
\caption{Histograms of skewness values for stars fainter (dashed
line) and brighter (solid line) than $B \approx 13.5$. The histogram 
for the fainter stars is broader than the one for brighter stars }
\label{fig16}
\end{figure}
       
There are two other facts which reinforce the conclusion that asymmetry has
no influence on the data:

1) The largest skewness values correspond to the fainter images, which
are much sensitive to plate defects and noise which might alter their
density profile. Figure~\ref{fig16} shows the histograms for the skewnesses of 
stars with magnitude $B \ga 13.5 $ and $B \la 13.5 $.
The standard deviation of both histograms has been calculated and the results
show that the distribution for the brightest stars around 0 is narrower
than the corresponding one for the faintest ones, as one can see from 
Table~\ref{table33}. At the bright end the non-linear response of the plate 
also
contributes to this effect as the stars are increasingly broader thus masking
features visible in fainter star images.

\begin{table}
\caption[]{Distribution of skewness values around zero}
\begin{flushleft}
\begin{tabular}{lllll}
\hline\noalign{\smallskip}
$B$ magnitude $^{a}$& data $^{b}$ & $\sigma_1$ $^{c}$ & $\sigma_2$ $^{d}$ & 
$\sigma_3$ $^{e}$ \\ 
\noalign{\smallskip}
\hline\noalign{\smallskip}
$ \la 13.5$ & 337 & $0.16$ &  $0.16$ & $0.15$ \\
$ \ga 13.5$ & 910 & $0.25$ &  $0.30$ & $0.27$ \\
\noalign{\smallskip}
\hline
\end{tabular}
\footnotesize{\ \ \ \ \ \ \ \ \ \ \ \ \ \ \ \ \ \ \ \ \ \ \ $^{a}$ magnitude interval considered \\ $^{b}$ number of stars used 
in each case \\ $^{c,d,e}$ standard deviations of the distributions of skewness 
values computed for exposures 1, 2 and 3, respectively}
\end{flushleft}
\label{table33}
\end{table}

This suggests that the images are symmetric and asymmetries only arise 
because of noise in faint stars. This is a normal effect in photographic 
plates and it is not a specific problem of the CdC plates.

\begin{figure}
\def\epsfsize#1#2{.65\hsize}
\centerline{\epsffile{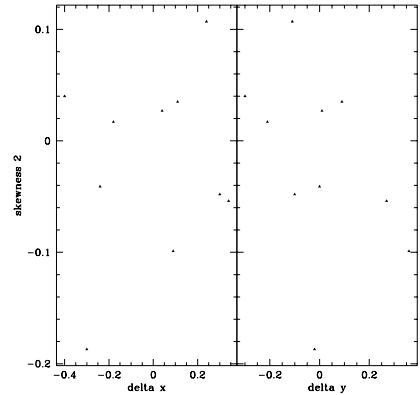}}
\caption[]{Residuals for star positions after the plate reduction with an
external catalogue vs. skewness values. Residuals are independent of the 
skewness. As an example we show here the skewness computed for exposure 2. 
Analog diagrams are obtained for skewness values of exposures 1 and 3}
\label{fig17}
\end{figure}

2) The residuals for positions on the plate after the plate reduction with the 
catalogue (PPM) are not a function of the skewness values (Fig.~\ref{fig17}).

As a conclusion one can say that at least the asymmetry in the images is not 
important enough to have a detectable influence on the data. Anyway, it has
been found that optical aberrations in CdC plates can be very complicated when
the centres of the telescope lenses were misaligned, being dependent on 
magnitude and colour (Zacharias, private communication). And, as usual when 
working with photographic plates, plate borders are of lower quality than the 
plate central region as they introduce some trends in the final position 
residuals when performing the plate reduction.

\section{Astrometric reduction}
\begin{figure}
\def\epsfsize#1#2{.35\hsize}
\centerline{\epsffile{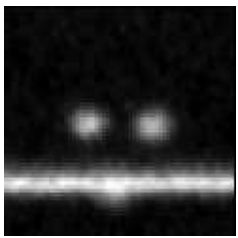}}
\caption{Example of a triple star image of which one is on a grid line}
\label{fig6}
\end{figure}

\begin{figure}
\def\epsfsize#1#2{.65\hsize}
\centerline{\epsffile{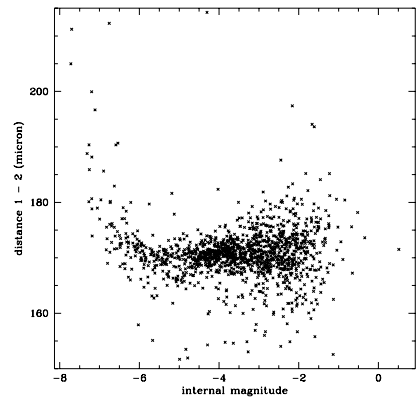}}
\caption{Distance between exposures 1 and 2 of each star as a function of a 
measure of the star brightness. The apparent repulsion between exposures 1 
and 2 due to the Kostinsky effect is more evident for the brightest stars}
\label{fig3}
\end{figure}

Further plate characteristics influencing the final astrometric accuracy 
have been found. The presence of a grid in the plates prevented us from 
considering stars lying on or very close to the lines (Fig.~\ref{fig6}).
Because of this we loose approximately $15\%$ of the total number of stars in 
the plate. These lines introduce changes in the density of the images and in 
the positions of the nearby ones due to the Kostinsky effect 
(Kostinsky 1907, Ross 1921). This effect shows up also when images of stars lie
very close to each other. 
The developer is exhausted in the zones where two images are in contact 
or very close and the amount of the products from the development process 
is larger there than in other zones.
These products inhibit the chemical reaction which takes place during the
development and the final result is an apparent repulsion of the images.
Due to the small separation between
the three exposures ($\approx 170 \mu$m), they begin to merge in the case of
 stars of magnitude $B\la 12$, and consequently, the Kostinsky effect
introduces changes in the ``true'' positions of the images. These can
be quite large: changes of $\approx 40 \mu$m have been 
detected (see Fig.~\ref{fig3}). This effect is very difficult to model: it 
changes from plate to plate as it depends on the development process 
of each plate individually. The determination of its variation as a function 
of the stellar magnitude is one of the most important targets in order to be 
able to correct for it.

\subsection{Plate-to-plate transformation}
To obtain the internal accuracy of the plate, we have transformed the 
positions of two of the three exposures to the reference system of the
third one by means of a linear transformation, and the deviations from
the new computed positions were obtained. It is equivalent to consider
three different single exposure plates. The linear transformation was
computed by least squares fitting of the positions of 552 stars 
in the field whose magnitudes ranged from $B=12$ to $B=14.5$ to avoid
large errors introduced by very bright or very faint stars.
Assuming that all exposures
equally contribute to the deviations, the rms of these deviations can be
calculated for each individual exposure using 1108 stars with
$B \leq 15.5$ (thereby excluding the ten ones with larger errors). Since 
these plates were taken originally to be complete up to the 
$14^{\mbox{\footnotesize th}}$ magnitude, images fainter than 
$B \approx 15.5$ have poor quality. Almost all the stars with larger errors 
were fainter than $B=13$. The results are presented in the first row in 
Table~\ref{table1}.

\begin{table}
\caption[]{rms of the position resulting 
from the \sl internal \rm plate comparison}
\begin{flushleft}
\begin{tabular}{lllll}
\hline\noalign{\smallskip}
field $^{a}$ & model  $^{b}$& $\sigma(\Delta x)$  $^{c}$
& $\sigma(\Delta y)$  $^{d}$& no. of stars  $^{e}$\\
\noalign{\smallskip}
\hline\noalign{\smallskip}
whole plate          & 22 & $\pm 0^{''}\!\!.16$ &$ \pm 0^{''}\!\!.16$ & 1108\\
$1^{o} \times 1^{o}$ & 22 & $ \pm 0^{''}\!\!.12$ & $ \pm 0^{''}\!\!.08$ & 394\\
plate centre         &  22 & $\pm 0^{''}\!\!.09$  & $\pm 0^{''}\!\!.09$ & 648\\
whole plate          & 12 & $\pm 0^{''}\!\!.15$ & $\pm 0^{''}\!\!.12$ & 1108\\
plate centre         &  12 & $\pm 0^{''}\!\!.10$ & $\pm 0^{''}\!\!.10$ & 648\\
\noalign{\smallskip}
\hline
\end{tabular}
\footnotesize{$^a$ zone of the plate used for analysis\\ $^b$ number of free 
parameters in the model to determine the positions \\ $^{c,d}$
rms of the position in $x$,$y$, respectively\\ $^e$ number of stars used}
\end{flushleft}
\label{table1}
\end{table}

This analysis was also performed over the smaller region of 
$1^{\mbox{\scriptsize o}} \times 1^{\mbox{\scriptsize o}} $ in the plate 
covering the zone in which the cluster M\,67 is located, since a good external
catalogue is available for this area (Girard et al. 1989).
The linear model was in this case computed using a total of 215 stars
with magnitudes ranging from $B=12$ to $B=14.5$. The final rms were
obtained from a set of 394 stars brighter than 15.5 (second row in 
Table~\ref{table1}).
One can see that $\sigma(\Delta x)$ are larger than those for
$\sigma(\Delta y)$. This may indicate that there are some kinds of local 
distortions which show up when studying only a small plate region, while they 
are absorbed by the model when considering the whole plate.
They can be due to slight emulsion displacements during the plate drying 
process or because storage of the plate in vertical position for almost 
one hundred years, to small errors in the telescope tracking, or to 
optical distortions as the small field considered around M\,67 is at the plate 
edge, where these distortions are expected to be stronger. Also the 
background, and consequently the noise, was found to be larger at the 
plate borders (see Fig.~\ref{fig24}).

To determine if these ``plate border effects'' could account for the fact that
 $\sigma(\Delta x) > \sigma(\Delta y)$ an internal reduction using stars 
closer than 5 cm to the plate centre was carried out. Accuracies of
$\sigma(\Delta x)=\sigma(\Delta y) = 0^{''}\!\!.09$ were found. This result 
clearly reflects the poorer quality of plate borders. Therefore, it seems 
that the triple Gaussian model can deal with this trend of $\sigma(\Delta x) > 
\sigma(\Delta y)$ when considering the whole plate, although the final 
accuracies are poorer than when only considering the plate central region.

\subsection{Plate-to-catalogue reduction}
\label{sec2}
For the $1^{\mbox{\scriptsize o}} \times 1^{\mbox{\scriptsize o}}$ region mentioned before containing the 
cluster a separate reduction was performed. The external comparison was 
carried out with an astrometric catalogue specifically built for this 
field (Girard et al. 1989). It has an external error of $0^{''}\!\!.16$ in 
the star positions for 1950.8, the catalogue weighted mean epoch. The cluster is
located $10^{'}$ north of the field centre. The plate model used consisted 
of polynomials in $x$ and $y$ up to the second order and the reduction was 
performed using the mean position over the three positions corresponding 
to the three exposures we have per star. In this way it is possible to avoid to 
some degree the consequences of the Kostinsky effect over the brightest stars.
The rms of the deviations are shown in the second row in Table~\ref{table11}. 
In these calculations three stars with exceptionally large deviations were 
not taken into account and the star showing the largest deviation after the 
reduction was also rejected.

\begin{table}
\caption[]{rms of the triplet mean position resulting from the plate 
reduction with an \sl external \rm catalog}
\begin{flushleft}
\begin{tabular}{lllll}
\hline\noalign{\smallskip}
catalog $^{a}$ & model $^{b}$ & $\sigma(\Delta x)_m$ $^{c}$ 
& $\sigma(\Delta y)_m$ $^{d}$ & no. of stars $^{e}$\\
\noalign{\smallskip}
\hline\noalign{\smallskip}
PPM    & 22 & $0^{''}\!\!.23$ & $0^{''}\!\!.19$ & 10\\
Girard & 22 & $0^{''}\!\!.20$ & $0^{''}\!\!.20$ & 196\\
PPM    & 12 & $0^{''}\!\!.16$ & $0^{''}\!\!.13$ & 10\\
\noalign{\smallskip}
\hline
\end{tabular}
\footnotesize{$^{a}$ catalog employed \\ $^{b}$ number of free parameters in the
 model used to determine the star positions\\ $^{c,d}$ rms of each triplet mean 
$x$,$y$ coordinates \\ $^{e}$ number of stars used}
\end{flushleft}
\label{table11}
\end{table}

The reduction of the full plate was performed using the PPM as an external
catalogue. Only 10 stars (with magnitudes $B=8.24 \leq B \leq 11.13$)
could be used in the reduction as in the PPM stars are normally brighter than
$V=11$. The triple images of these stars are quite blended in our plates and 
the errors for the parameters obtained in their Gaussian fitting are a bit worse
than the ones corresponding to fainter stars. The plate model used consisted of 
polynomials in $x$ and $y$ up to the second order and the rms of the deviations 
are shown in the first row in Table~\ref{table11}. The larger values for 
$\sigma(\Delta x)$ than for $\sigma(\Delta y)$ can be due to the fact that 
the reference stars happen to be in regions of the plate where the optical 
distortions are stronger along $x$-axis than $y$-axis and the fitting algorithm 
is slightly poorer in the case of the brightest stars because when the 
exposures merge the determination of the exposure centres is less accurate. 
Errors during the telescope tracking can also be an explanation.

\begin{figure}
\def\epsfsize#1#2{.65\hsize}
\centerline{\epsffile{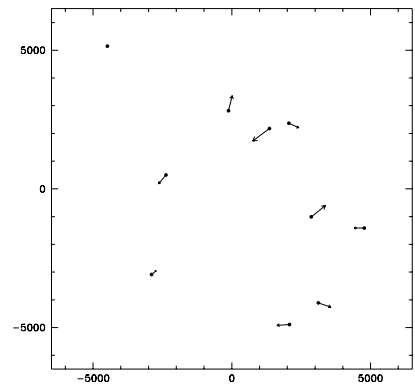}}
\caption[]{Residuals for the mean position of the three exposures after
the plate reduction with the PPM catalog as a function of the plate position. 
Arrow lengths have been enlarged 1500 times}
\label{fig11}
\end{figure}

In Fig.~\ref{fig11} the residuals in the mean position are plotted as a 
function of the plate position. They are uncorrelated so it seems that there 
are no global systematic errors present in the plate although ten reference 
stars are clearly too few to categorically affirm that there are no systematic
errors left.

\section{Photometry}
\label{photometry}

An internal instrumental magnitude was computed from the
measured density of each exposure. For this purpose,
the volume under each Gaussian was considered and an internal instrumental
magnitude was defined as

\begin{equation}
m_{k} = - 2.5 \log (v_{k}),
\end{equation}

\noindent
where $v_{k}= A_{k} ( \pi a_{k} b_{k})$, $A_{k}$ being the peak density of
the $k$-th exposure in the triple image, and $a_{k},b_{k}$ the semiaxes of
the elliptical contour. But this is true only in the case of non-saturated 
images for which a Gaussian model is correct.

For saturated images it happens that the volume of the fitted Gaussian 
is smaller than the one which would correspond to a linear detector 
without saturation because the image is, in some way, cut by the 
saturation limit of the plate.

Let us suppose that we have a Gaussian
and that we cut it in two parts at a certain height $h$. The ratio between the
volume of the ``lower'' part, $V$, and the total one, $V_{\mbox{\footnotesize c}}$, is 
found to be:

\begin{equation}
\frac{V}{V_{\mbox{\footnotesize c}}}=h \times (1-\ln h) $$
\end{equation}

\noindent
where $h$ is the truncation height and the total height is assumed to be 1.

This allows to obtain the ``true'' Gaussian volume $V_{\mbox{\footnotesize c}}$ in the
case of saturated (truncated) images as a function of the plate saturation
peak (truncation ) $p_{\mbox{\footnotesize s}}$, the image area $a$, and the measured
volume $V$:

\begin{equation}
V_{\mbox{\footnotesize c}}=a \times p_{\mbox{\footnotesize s}} \times \exp\{\frac{V}{a \times p_{\mbox{\footnotesize s}}}-1\} $$
\end{equation}

This $V_{\mbox{\footnotesize c}}$ is an estimation of the volume which can be 
used in the evaluation of the 
internal magnitude when dealing with saturated images. The measured 
volume $V$ is a Gaussian one, corresponding to the volume under a bivariate 
Gaussian: It is the semimajor times the semiminor image axis times maximum 
amplitude of image peak density. We are interested also in comparing the 
magnitudes estimated from $V$ (uncorrected) and the ones from
$V_{\mbox{\footnotesize c}}$ (corrected).
                                                          
\begin{figure}
\def\epsfsize#1#2{.75\hsize}
\centerline{\epsffile{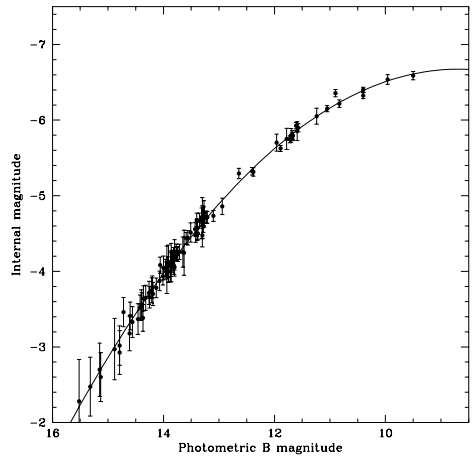}}
\caption[]{Magnitude calibration using the measured internal magnitudes
and a photometric sequence}
\label{fig20}
\end{figure}

The photometric reduction was performed with 102~stars from the SIMBAD 
data base which have been identified in the plate. More than $70\%$ of these
data comes from a work by Eggen \& Sandage (1964), which ensures that we
are dealing with a mostly homogeneous sample.

A second order polynomial was used to fit the data and
accuracies of $\sigma=\pm 0^{\mbox{\scriptsize m}}\!\!.09$ (case of 
uncorrected magnitudes) and 
$\sigma_{\mbox{\footnotesize c}}=\pm 0^{\mbox{\scriptsize m}}\!\!.10$ (when
corrected magnitudes are introduced) were
obtained. A total of two outliers and seven poor quality stars were removed 
from the sample prior to the fitting. An analysis of these poor quality stars
revealed that six of them are the brightest stars in the sample located at a
distance from the plate center larger than $4.7$ cm, where optical aberrations
produce important distortions on the images. The seventh one removed is the
brightest star in the whole sample. As the method of  corrected volumes relies
on the proper estimation of the image area, it is very sensitive to distorted
cases such as the ones just described. Thus, a wrong estimation of the image
area produces a wrong estimation of the correction to the image volume and this
method cannot deal with these pathological cases.
The method of corrected magnitudes constitutes an experimental model
which forces the calibration curve to be closer to linear at the bright end
and may be used if such a property were essential. In
Figure~\ref{fig20} we show how the fitting looks like when using uncorrected
magnitudes, after dropping off the seven problematic cases. The error bars in 
this figure are the propagated errors of the profile parameter errors.
                                             
One can compare the accuracy obtained here, 
$\sigma=\pm 0^{\mbox{\scriptsize m}}\!\!.09$ with the accuracy in 
Eggen \& Sandage paper, which is 
$\sigma_{\scriptsize{\mbox{ES}}}~\approx~\pm 0^{\mbox{\scriptsize m}}\!\!.02$. It is
clear that $\sigma_{\scriptsize{\mbox{ES}}}^{2}$ is negligible in comparison 
with $\sigma^{2}$, therefore one can state that the main sources of the found 
rms error $\sigma$  must be related to the quality of our CdC plate and to the
specific reduction method described here.
          
\section{Reducing the number of free parameters in the model}
\label{pca}

A problem present in the three-gaussian model fitted to the data is the
large number of free parameters, 22 in total per triple image. It is clear
that some of these parameters must be correlated as the three exposures
were intended to be as similar as possible. Therefore a model
with smaller number of degrees of freedom would be interesting to consider
because:

a) The fitting process would be more robust with a smaller number of
free parameters

b) The fitting algorithm would be faster

Principal Components Analysis (PCA) (Brosche 1973; Whitney 1983, for example)
is a method which first obtains a new set of uncorrelated variables by 
determining the set of eigenvectors of the original data correlation matrix.
The corresponding eigenvalues give the portion of total variation
in the original data for which each eigenvector accounts for. Then the
eigenvalues are sorted and split into two groups: The ones with larger
values putatively represent the signal, and 
the remaining ones represent the noise. The programme used 
(Lentes 1985) contains a modification of the correlation matrix in order to 
take care for error covariance. 

We found that it is possible to reduce the number of free parameters to not 
less than 7. 
This minimum is defined by the number of eigenvalues with larger values 
representing the
actual data.

\begin{figure}
\def\epsfsize#1#2{.65\hsize}
\centerline{\epsffile{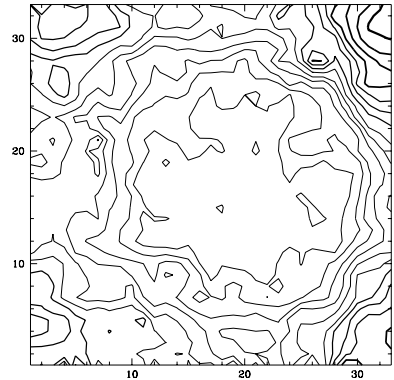}}
\caption[]{Contour plot of different background levels. Thicker lines 
correspond to larger values.The non-uniformity of the background across the 
plate is obvious, as well as the large inhomogeneity toward the plate edges}
\label{fig24}
\end{figure}

A question can be posed now: Is there any set of less than 22 parameters which 
seems suitable {\sl a priori} to represent the data? Some characteristics 
in our problem 
can help us to find a possible candidate. For example, the three exposures 
should in 
principle be identical, as they were supposed to be given the same exposure 
time in each 
individual plate and were taken under approximately the same 
atmospheric conditions, as there was only about half an hour between 
consecutive 
exposures. More explicitly, we have found that, in average, in the current 
data the values of the parameters in an individual exposure are compatible with the case 
of being equal to the analogue ones in the other two exposures for each single star.
This means that it is basically correct to use only one peak density value $A$ for the 
three exposures, and the same holds for $\sigma_x$, $\sigma_y$, $s$ and $t$. The only 
parameter left, the background, was found to be varying in a not at all regular way over 
the plate (Fig.~\ref{fig24}). However, the scale of this variation is much larger than 
just the $0.81 \times 0.81$ mm windows containing the triple images and no errors are 
introduced when taking it to be constant. Thus, it was considered to be the same inside 
the whole frame containing the triple image

The reasonings above, thus, lead in total to a set of 12 free parameters, that is, the 
6 mentioned just before and 6 position coordinates for the centres of the three 
exposures. These have been the degrees of freedom of the new model applied in 
Sect.~\ref{sec1}.

The question remains whether it is possible to only have 7 degrees of freedom 
without loosing an important fraction of the original information. First, one 
would expect a strong correlation between $\sigma_x$ and $\sigma_y$ (and 
it is so indeed) hence one of them can be computed as a function of the 
other one. One has to pay attention to the fact that because of the presence
of field curvature this function will depend on the distance to the plate
centre.

Moreover, could it be possible to know the position of each
exposure only knowing the triangle central coordinates? For this to be
correct the triangle geometry should be constant through the
plate. The angles which the position vectors of each exposure in a triplet
form with an $x$-axis parallel to the $x$ plate scan axis and as origin the
triangles geometric centre have been calculated, together with their rms. Its
dependence on the position in the plate and on the distance to the plate
centre was studied and it was found that exposure 1 suffers a displacement when
approaching one of the plate corners. It also happens to exposure 3, but
to a smaller degree. The total dispersion of the angle values is equivalent
to an image displacement along an arc of length of $0^{''}\!\!.12$, 
$0^{''}\!\!.03$ and $0^{''}\!\!.005$ for exposures 1, 2 and 3, respectively. 
One can compare this
with the residuals when performing the comparison with an external catalogue
(Table~\ref{table11}). If the total error in the image position is taken to 
be $\Delta d=\sqrt{(\sigma(\Delta x))^{2} + (\sigma(\Delta y))^{2}}$, we obtain
$\Delta d=0^{''}\!\!.30$, which is larger than $0^{''}\!\!.12$ . Thus, angle 
dispersion is not the main source of error in the position in this case. We 
have also checked that the ratios among the triangle sides are not dependent 
on the magnitude.

A model for the three exposure positions as a function only of the geometrical
triangle centre position in the plate is then possible and this
would lead to a set of 7 free parameters, as predicted by the PCA: $A$, 
$\sigma_x$ (or $\sigma_y$), $s$, $t$, $x_c$, $y_c$ and the background ($B$).

\section{Reduction with a 12 free parameter model}
\label{sec1}

A new reduction was then performed, modeling the data now with a 12
parameter model instead of the 22 parameter one used before. This new
model was simply the addition of three Gaussian curves with identical
$A$, $\sigma_x$, $\sigma_y$, $s$, $t$ and $B$, together with 
the corresponding 6 coordinates for the three image centres:

\begin{eqnarray}
\begin{array}{l l}
D_{ij}= & B + \sum\limits_{k=1}^{3} A \exp \left\{ - \frac{1}{2} \left[ \frac{1}
{1-t^{2}} \left( \left( \frac{x_{ij}- x_{ck}}{\sigma_{x}} \right)^{2} \right. 
\right. \right. \\ 
& 
\left. \left. \left. + \left( \frac{y_{ij}-y_{ck}}{\sigma_{y}} \right)^{2} 
-  2t \left( \frac{x_{ij}-x_{ck}}
{\sigma_{x}} \right) \left( \frac{y_{ij}-y_{ck}}{\sigma_{y}} \right) 
\right) \right]^{s} \right\}
\end{array}
\end{eqnarray}

\subsection{Plate-to-plate transformation}

An internal comparison of the plate was again performed following the same
method already described. There is an improvement in comparison with the 
results obtained when using the 22 free parameter model, as can be seen
in the fourth row in Table~\ref{table1}. We noted that 
$\sigma(\Delta x)~>~\sigma(\Delta y)$, again due to border effects as 
we obtained 
$\sigma(\Delta x)~=~\sigma(\Delta y)~=~0^{''}\!\!.10 $ when working with a
circular area of radius 5 cm around the plate centre (note that it is
the same obtained for the same region with the 22 parameter model). 
It seems from Table~\ref{table1} that the 22 free parameter model was able 
to deal with this trend when considering the whole plate while the 12 
parameter model was not. This means that distortions in the plate borders
introduce significant deviations from the hypothetical identity of the three
exposures, which is assumed in the 12 parameter model.

\subsection{Plate-to-catalogue reduction}

A remarkable improvement is obtained when working with the 
12 parameter model when performing the reduction with the PPM as before
(see last row in Table~\ref{table11}). Again, 
$\sigma(\Delta x)~>~\sigma(\Delta y)$ due to the poor quality of the 
reference stars and 
possible telescope tracking errors, as explained in Sect. \ref{sec2}.

\subsection{Photometry}
A second order polynomial has been fitted to the data as before, after removing
from the sample the seven ones which were found to be to severely affected
by saturation and optical aberrations (see Sect. \ref{photometry}). The 
magnitudes are obtained with an accuracy of 
$\pm 0^{\mbox{\scriptsize m}}\!\!.10$, to be compared with the one obtained
with the 22 parameter model, 
$\pm 0^{\mbox{\scriptsize m}}\!\!.09$ .
It is clear that both methods lead to the same accuracy in the photometry.

\section{Conclusions}

We showed that it is possible to make use of the information 
stored in the CdC plates in spite of the large variety of problems 
that they exhibit (triple exposures, optical aberrations, grid lines, 
emulsion saturation). 

A general procedure in their analysis has been developed, based in the
fitting to these triple exposures of a model consisting on the sum of 
three bi-variate Gaussian distributions in arbitrary position.

A detailed analysis of some problems in the plates was carried out. It was
found that optical aberrations which result in symmetrically elongated images
(spherical aberration and field curvature) dominate  others which 
could be also present and would introduce asymmetries in the images (coma).
We have seen that these particular aberrations have no influence in the 
final astrometric accuracy as the bivariate Gaussians are able to correctly 
model these elliptical images present mainly in the plate corners.

The internal plate accuracies obtained per exposure
are $\sigma(\Delta x)~=~\pm 0^{''}\!\!.15$ and 
$\sigma(\Delta y)~=~\pm 0^{''}\!\!.12$ . Improvement is obtained when 
considering only a 
circular area around the plate centre as in this case accuracies were 
$\sigma(\Delta x)~=~\sigma(\Delta y)=\pm 0^{''}\!\!.10$ for both the 22 and 
12 parameter models. This demonstrates the poorer quality of images at the 
plate borders due to optical distortions mainly.

The reduction with an external catalogue (PPM) gives values of  
$\sigma(\Delta x)~=~\pm 0^{''}\!\!.16$ and 
$\sigma(\Delta y)~=~\pm 0^{''}\!\!.13$ . This reduction was 
done with problematic data as all the stars considered were among the 
brightest ones on the plate and they exhibited larger errors in the fitting 
due to the blending of the three exposures. Only 10 stars could be used as 
reference stars. The differences among these $\sigma(\Delta x)$ and 
$\sigma(\Delta y)$ are most probably due to the poorer quality of the bright 
images and/or problems in the telescope tracking. 

The 12 free parameter model is found to be equivalent to the 22 parameter
model for the internal reduction but for the external reduction results are 
better with the first one. This is because although the 22 parameter model is
capable of reproducing the data in a more detailed way it is also more 
sensitive to individual image errors.

A photometric reduction was also performed using a homogeneous sample of
data from Eggen \& Sandage (1964) mainly, with a resulting accuracy of 
$0^{\mbox{\scriptsize m}}\!\!.09$. Photometry with the 12 parameter model 
gives the same result, so the photometry is not affected by the change in 
the model in these two particular cases.

Improvements to the algorithm have been also outlined on the basis of the
results obtained with a Principal Component Analysis of the data which
indicates that a model with at least 7 free parameters instead of 22 is 
possible. These parameters are: one peak density ($A$), one profile width 
$\sigma$, one saturation parameter ($s$), one orientation parameter ($t$), 
one background ($B$) and the triangle centre coordinates ($x_c, y_c$).

\begin{acknowledgements}

The authors wish to thank Drs. H.-J. Tucholke and M. Geffert
for providing us with part of the software used in this work, together with 
Dr. M. Odenkirchen, Dr. A. Fern\'andez-Soto and Prof. Dr. K. S. de Boer for 
very useful discussions and 
comments. Dr. Th. Lentes provided us with his software to perform the 
Principal Components Analysis. We are very grateful to Bordeaux Observatory 
for providing us several CdC plates from their archives. We also 
thank SIMBAD Data Base at the CDS (Strasbourg, France) for providing us 
part of the data used.

This work has been carried out in the frame of a European Community (EC) 
Network entitled ``Salvaging an astrometric treasure'' (project no. 
CHRX-CT94-0533) within the ``Human and Capital Mobility'' program. The 
authors duly acknowledge the support of the EC, specially A. Ortiz-Gil and M. 
Hiesgen with regards to their fellowships.
\end{acknowledgements}

\end{document}